\documentclass{article}

\usepackage[english]{babel}

\usepackage[letterpaper,top=2cm,bottom=2cm,left=3cm,right=3cm,marginparwidth=1.75cm]{geometry}

\usepackage{amsmath}
\usepackage{graphicx}
\usepackage[colorlinks=true, allcolors=blue]{hyperref}

\title{Alignment of a Total Automation Economy}
\author{David McAllester}

\begin{document}
\maketitle

{\large
\centerline{Consumption is the end purpose of production.}

\vspace{1ex}
\hfill \text{Adam Smith\hspace{5 em} ~}

\vspace{3ex}

\centerline{History is driven by economics.}

\vspace{1ex}
\hfill \text{Historical Materialism --- Karl Marx \hspace{5 em}}

\vspace{3ex}
\center{Government of the people, by the people, and for the people, must not perish.}

\vspace{2ex}
\hfill \text{Abraham Lincoln, The Gettysburg Address \hspace{5 em}}
}

\vspace{3ex}
\begin{abstract}
We consider economic theory from the perspective of a total automation economy, one with no human involvement in production either in manufacturing or in management. One can naturally ask whether a total automation economy is fundamentally a centrally planned economy or, alternatively, whether efficiency demands decentralization
into local decisions by competing agents --- agentic production.  A soviet economist, Leonid Kantorovich, developed linear programming as a method companies or governments can use to optimize production.  Ironically, he is also generally credited with showing that the most efficient production is achieved through decentralization --- a free market economy with competing agents. Here we review Kantorovich's dualization in detail. We take the objective of the economy to be maximizing production weighted by (human) market price.  A fundamental issue is whether an automated pursuit of this objective might have alignment vulnerabilities as the economy evolves.  Another question is whether dualization provides insight into the utility of agentic AI systems (multi-agent AI systems) generally.
\end{abstract}

\section{Introduction}
Here we consider Kantorovich's dualization theorem~\cite{kantorovich1939} from the perspective of a total automation economy --- one where no human is used in the production of goods and services. The statement of the theorem assumes that the goal of industrial production is maximizing consumption weighted by market price. One can think of market pricing as a form of democracy where people ``vote" on what is to be produced simply by purchasing what they want. The ability of the economy to respond to the purchases of consumers can be viewed as a form of alignment of the economy with human value.

There are valid fairness issues to be address in treating purchases as reflective of social welfare.  Obviously rich people have more "voting power" (purchasing power) than poor people.  The needs of poor people are more pressing and should perhaps be given even more weight than the needs of rich people (instead of the reverse). Also, people in different states of health have different significance of their needs. A final issue is that different people simply have different fundamental values perhaps stemming from differences in their fundamental natures. Market pricing would seem to better reflect human value if we assume that all consumers have the same universal basic income.  However, here we put the fairness issues aside and simply assume a mathematical model in which the objective is to maximize consumption weighted by price.

In this brief paper we review Kantorovich's dualization theorem and informally consider some issues related to it's application. Although we present only a static optimality condition in a linear model of production processes, in practice the economy would evolve due to advances in technology and the production processes would be nonlinear. Control of the economy would presumably be based on some form of gradient ascent on the objective.  It seems that attention must be paid to the existence of any possible alignment vulnerabilities.  Another issue is whether dualization of an objective function provides understanding of the general value of agentic AI systems --- problem solving using a variety of differently motivated agents.

\section{Formal Set Up}

\newcommand{\inp}{\mathrm{In}}
\newcommand{\out}{\mathrm{Out}}
\newcommand{\en}{\mathrm{Endow}}
\newcommand{\net}{x}
\newcommand{\good}{\mathrm{Welfare}}

The formal set up here is due to Kantorovich~\cite{kantorovich1939}, who won a Nobel Prize in economics (1975) for formulating and analyzing this model.
We assume a set of $N$ commodities indexed by $i$. We assume $K$ processes indexed by $k$ and let  $z_k$ be a rate at which process $k$ runs. Each process consumes commodities as inputs and produces commodities as outputs.  For commodity $i$ and process $k$ we write $\inp_{k,i}$ for the rate commodity $i$ is consumed by process $k$ when process $k$ runs at a unit rate.   Similarly $\out_{k,i}$ denotes the rate of production of $i$ when $k$ is run at unit rate.  We assume an industrial endowment which is a supply of natural resources (for example land) and let $\en_{i}$ be a rate at which commodity $i$ is produced as an endowment. Given a set of processes running at given rates $z_k$ we have a net production of rate $\net_i$ for commodity $i$ given by
$$\net_i(z) = \en_i + \sum_k z_k(\out_{k,i} - \inp_{k,i})$$
We say that a set of rates $z_k \geq 0$ is feasible if for each commodity $i$ we have $\net_i(z) \geq 0$. We cannot consume more of commodity $i$ in production than is made available by the endowment plus production.
A commodity $i$ with $\net_i(z) = 0$ will be called an internal commodity --- internal commodities are used in production but are not provided to consumers. Engines might be used in making cars but never offered to consumers.
A commodity will be called external if it is not internal (and hence offered to consumers in a market).

We define the purpose of the the economy to be that of maximizing consumption weighted by price.
To define prices we invoke the Arrow-Debreu theorem~\cite{arrowdebreu1954} which implies that for a given net production $\net_1,\ldots,\net_n$
there exists a market equalibrium price vector $p_1,\ldots,p_n$ such that when human consumers are offered the commodities in quantity $\net_1,\ldots,\net_n$ at prices $p_1,\ldots,p_n$ the market clears --- consumers buy exactly the amount produced.
A fundamental issue is that $p_i$ is not defined for internal commodities which are not offered to consumers.
This is a wrinkle in the optimization problem that is not present in the general analysis of KKT conditions
for constrained optimization. However, as explained below, we adopt the convention that $p_i = 0$ for any internal commodity $i$.

\section{A Central Planning Optimality Condition}

We will work with differential changes in production rates $d z_1,\ldots d z_k$.  Differentiating the definition
of $\net_i(z)$ we have
$$d \net_i = \sum_k(\out_{k,i} - \inp_{k,i})dz_k$$
A process $k$ will be called inactive if $z_k = 0$ and called active otherwise.
For feasible production rates $z$ we say that a differential update direction $dz_1,\ldots,dz_K$ is feasible
if for each inactice process $z_k$ we have $dz_k \geq 0$ and
for each internal commodity $i$ we have $d\net_i \geq 0$. A feasible update direction
must not violate the feasibility restrictions on $z$. However a feasible update direction might convert an internal commodity to an external commodity or vice-versa or convert an inactive process to an active process or vice-versa.

\begin{quotation}
\noindent {\bf Optimality}:
We say that a feasible setting for production rates $z_1,\ldots,z_k$ is (first order) optimal if for any feasible
update direction $dz_1,\ldots,d z_K$ we have $\sum_i p_i d\net_i \leq 0$.
\end{quotation}

If this optimality condition fails then there exists an update condition
with positive gradient of the objective.  If this first order optimality condition holds
then we might still be at saddle points or even local minima. There might also be a differential update direction that converts an internal commodity to an external commodity and improves the objective of converting
an an internal commodity $i$, with $p_i= 0$, to an external commodity with a strictly positive market value.

In practice the economy would evolve by some form of gradient ascent on the production rates $z_k$.  Note that the market prices are allowed to be nonlinear in the production rates $z_k$ and therefore must be measured repeatedly during any gradient ascent process.

\section{Agents Arising from Duality}

We now apply the KKT (Karush-Kuhn-Tucker) conditions on constrained optimization~\cite{kuhntucker1951,karush1939}.  We take the objective function to be
$\sum_i p_i \net_i$ with $p_i$ defined on all commodities $i$ but set to zero on internal commodities.
Applying the KKT conditions we introduce a Lagrange multiplier $\lambda_i \geq 0$ for each constraint $\net_i \geq 0$
and a Lagrange multiplier $\mu_k \geq 0$ for each constraint $z_k \geq 0$.  We set $\lambda_i = 0$ on external commodities and $\mu_k = 0$ on active processes.

We have that only one of $p_i$ and $\lambda_i$ can be non-zero.  This is not just complementary slackness and is not a property of constrained optimization generally.  In the general case of constrained optimization the "price vector" $p$ is just the gradient of a fixed objective. Here $p$ can vary at different settings of $z$ as commodities switch from being internal to being external.  We have {\bf defined} $p_i = 0$ for internal commodities. The KKT conditions will
apply to the price vector as if that vector is a well-defined gradient of the objective.

Before applying the KKT conditions we first note that
$$\nabla_z\net_i(z) = \sum_k (\out_{k,i} - \inp_{k,i})\delta_k$$
Here we have that $\nabla_z\net_i(z)$ is a K-dimensional vector and
$(\out_{k,i} - \inp_{k,i})$ is the partial derivative of $x_i(z)$ with respect to $z_k$ --- a scalar.
This scalar becomes a $K$-dimensional vector when multiplied by $\delta_k$ --- the vector whose $k$th component is 1 and all other components are zero.

For maximizing $\sum_i p_i x_i$ with constraints of the form $x_i(z) \geq 0$ and $z_k \geq 0$ the KKT condition is
\begin{eqnarray*}
0 &= &\sum_i p_i \nabla_z \net_i(z) +  \sum_i \lambda_i \nabla_z x_i(z) + \sum_k \mu_k \delta_k\\
 & = & \sum_k \delta_k\left(\mu_k +\sum_{i}(p_i + \lambda_i)(\out_{k,i}-\inp_{k,i})\right)
 \end{eqnarray*}

The sum over $k$ is summing over orthogonal vectors so this equation yields that for each $k$ we have
$$\mu_k + \sum_i (p_i + \lambda_i)(\out_{k_i} - \inp_{k,i}) = 0$$
For every active process we have that $\mu_k = 0$ and hence all active processes must operating at zero profit
under the pricing $(p_i + \lambda_i)$.  For each inactive process we can satisfy the equation by setting $\mu_k$ provided that $(p_i + \lambda_i)(\out_{k,i}-\inp_{k,i}) \leq 0$. We now have the following localization theorem.

\begin{quotation}
\noindent {\bf Kantorovich Dualization:} A feasible setting of the process weights $z$ is first order optimal (as defined previously) if and only if there exists a "shadow price" $\lambda_i \geq 0$ for each internal commodity $i$ such that under the commodity pricing $p + \lambda$ each active process is operating at break-even (zero profit) and no inactive process is operating at a strictly positive profit.
\end{quotation}

The shadow prices $\lambda_i$ are Kantorovich's ``objectively determined valuations''~\cite{kantorovich1965} --- imputed prices for internal commodities supplied by the optimization itself rather than by any human consumption market.

\section{Summary}

A total automation economy run by AI should serve humanity by providing what people want as indicated by the purchases that they make. Kantorovich showed that a certain notion of centralized optimization of industrial production (setting prices to zero for internal commodities) is essentially equivalent to free market economics with agents pursuing profit.

Here we have discussed an optimality condition. In practice we must find some form a gradient ascent involving states of the economy with prices, shadow process and possibly nonzero profits and losses.  A fundamental design criterion of such a gradient ascent algorithm is that it remain aligned to the gradient of consumption weighted by price.

A independent question is under what conditions is the dual of an optimization problem usefully viewed as defining a collection of agents.  Might this explain the value of agentic AI more generally?

\section*{Acknowledgments}

I would like to thank Sergiy Verstyuk and Claude AI for useful comments on this paper.
\bibliographystyle{alpha}
\bibliography{sample}

\end{document}